\documentclass[12pt,preprint]{aastex}
\slugcomment{To appear in {\it The Astrophysical Journal}.}
\begin{document}
\title{Studies of the Relativistic Binary Pulsar PSR~B1534+12:
  I. Timing Analysis}
\author{I. H. Stairs}
\affil{National Radio Astronomy Observatory, Green Bank, West Virginia 24944}
\email{istairs@nrao.edu}
\author{S. E. Thorsett}
\affil{Department of Astronomy and Astrophysics, University
  of California, Santa Cruz, CA 95064}
\email{thorsett@ucolick.org}
\author{J. H. Taylor} 
\affil{Joseph Henry Laboratories and Physics Department,
       Princeton University, Princeton, NJ 08544}
\email{joe@pulsar.princeton.edu}
\and
\author{A. Wolszczan}
\affil{Department of Astronomy and Astrophysics, Pennsylvania State
University, University Park, PA 16802}
\email{alex@astro.psu.edu}

\begin{abstract}
We have continued our long term study of the double-neutron-star
binary pulsar PSR~B1534+12, using new instrumentation to make very
high precision measurements at the Arecibo Observatory.  We have
significantly improved our solution for the astrometric, spin, and
orbital parameters of the system, as well as for the five
``post-Keplerian'' orbital parameters that can be used to test
gravitation theory.  The results are in good agreement with the
predictions of general relativity.

With the assumption that general relativity is the correct theory of
gravity in the classical regime, our measurements allow us to
determine the masses of the pulsar and its companion neutron star with
high accuracy: $1.3332\pm0.0010M_\odot$ and $1.3452\pm0.0010M_\odot$,
respectively. The small but significant mass difference is difficult
to understand in most evolutionary models, as the pulsar is thought to
have been born first from a more massive progenitor star and then
undergone a period of mass accretion before the formation of the
second neutron star.

PSR~B1534+12 has also become a valuable probe of the local
interstellar medium.  We have now measured the pulsar distance to be
$1.02\pm0.05$~kpc, giving a mean electron density along this line of
sight of 0.011\,cm$^{-3}$.  We continue to measure a gradient in the
dispersion measure, though the rate of change is now slower than in
the first years after the pulsar's discovery.
\end{abstract}


\section{Introduction} \label{sec:intro}

The discovery \citep{ht75a} of the first pulsar in a binary system,
PSR~B1913+16, introduced the possibility of powerful new experimental
tests of gravity in the strong field and radiative regimes.  Over a
quarter century, measurements of this pulsar were shown to be in
excellent agreement with the predictions of general relativity
\citep{tw89,dt91,tay94}.  A dozen years ago the discovery \citep{wol91a}
of PSR~B1534+12---a bright, relatively nearby pulsar, also with a
neutron-star companion---promised the independent opportunity to
confirm the tests of gravity done earlier with PSR~B1913+16.
Furthermore, B1534+12's favorable, nearly edge-on orbital geometry
allowed new experimental tests that were complementary to those done
with PSR~B1913+16 \citep{twdw92}.

In earlier work \citep[hereafter Paper~I]{sac+98} we analyzed data
taken over a six year period with the Arecibo 305~m telescope, the
43-m telescope at Green Bank, West Virginia, and the 76-m Lovell
Telescope at Jodrell Bank Observatory, U.K.  We found that those data
were in agreement with the predictions of general relativity at the
limit of the measurement uncertainties: about 1\%.  We showed that the
orbital period was decreasing, as expected if the system is losing
energy in the form of gravitational radiation.  Because the observed
orbital decay rate is contaminated by kinematic effects which depend
on the unknown pulsar distance \citep{dt91,ctk94}, the radiation
predictions of general relativity cannot be tested with this pulsar to
better than $\sim30\%$ without an independent distance measurement.
However, with the assumption that general relativity is the correct
theory of gravity, we were able to invert the kinematic model to
measure the pulsar distance to 20\%.

Over the past four years, we have continued our studies of
PSR~B1534+12, using the recently upgraded Arecibo telescope together
with a new data acquisition system \citep{sst+00} that fully removes
the dispersive effects of the interstellar medium from the signal.  In
addition to extending the length of the data set by more than half,
these new data are of substantially higher quality than those
available for Paper~I.  In this paper, we report on timing studies
based on the complete 11.5-year Arecibo data set.  We present the
improved tests of general relativity that are now possible, give a
pulsar distance accurate to 5\%, and report on dispersion measure
variations.  In subsequent papers, we will describe the analysis of
secular variations in the pulsar radiation pattern that have been
interpreted as evidence for geodetic precession of the pulsar, and we
will discuss single-pulse studies.

\section{Observations} \label{sec:obs}
In this paper we discuss results of observations made with two
different observing systems at the Arecibo telescope.  Many of the
details have already been discussed in Paper~I, and are only
summarized here and in Table~\ref{tab:obsparms}.

Observations were made with the 305-m Arecibo telescope both before
its upgrade (1990--1994) and after its upgrade (1997--2002).  For our
purposes, the most important post-upgrade change was a significant
reduction in the system temperature at 1400\,MHz, leading to an
improvement in the signal-to-noise ratio achieved for a given
integration time and observing bandwidth.  During the period that the
Arecibo telescope was unavailable, observations were made with the
National Radio Astronomy Observatory (NRAO) 43-m telescope at Green
Bank, and with the 76-m Lovell telescope at Jodrell Bank Observatory,
U.K.  The latter two datasets were used in Paper~I, but, with the
abundance of high-quality post-upgrade Arecibo data, they contribute
nothing to the present solution and are therefore disregarded in this
work.

Two observing systems were used.  The Princeton Mark~III system
\citep{skn+92} is an ``incoherent'' system, in which the incoming
signal is decomposed into spectral channels with a filterbank.  The
effective time resolution is dominated by dispersion smearing of the
signal during its propagation through the ionized interstellar medium,
with some additional instrumental smoothing after detection of the
signal.  The signals in each frequency channel were folded at the
instantaneous topocentric pulsar period for three to five minutes,
then the channels were shifted to account for interstellar dispersion
and summed to produce a single total-intensity (summed-polarization)
pulse profile for each integration.

For the new post-upgrade data that we introduce here, we observed
with a ``coherent'' system: the Princeton Mark~IV instrument
\citep{sst+00}.  For each sense of circular polarization, a 5\,MHz
bandpass was mixed to baseband using local oscillators in phase
quadrature.  The four resulting signals were low-pass filtered at
2.35\,MHz, sampled at 5\,MHz with 4-bit quantization, and written to a
large disk array.  Some of the 1400\,MHz observations were taken with
10\,MHz of bandwidth and 2-bit quantization.  In the offline
processing, the undetected signal voltages were dedispersed using the
phase-coherent technique described by \citet{hr75}.  After amplitude
calibrations were applied, self- and cross-products of the right- and
left-handed complex voltages yielded the Stokes parameters of the
incoming signal.  These products were folded at the topocentric pulsar
period using 1024 phase bins, and pulse TOAs were determined from the
resulting total-intensity profiles.  A summary of the more important
parameters and statistics of all combinations of observing systems and
telescopes is presented in Table~\ref{tab:obsparms}.  Many of the
post-upgrade data come from observations on one or two days at
quasi-regular intervals; in addition we conducted 12-day intensive
``campaign-style'' observations in each of the summers of 1998, 1999,
2000 and 2001, with full orbital coverage at each epoch.  This mixture
of data permits a high-quality timing solution and allows us to
monitor, at each campaign epoch, the pulse shape changes due to
geodetic precession, which we will discuss in a subsequent paper.

As in Paper~I, we used the same TOA-fitting procedure for all data
sets.  Each observed profile was fitted to a standard template, using
a least-squares method in the Fourier transform domain \citep{tay92} to
measure its time offset.  The offset was added to the time of the
first sample of a period near the middle of the integration, thereby
yielding an effective pulse arrival time.  A different standard
template was used for each observing system and frequency; they were
made by averaging the available profiles over several hours or more.
Uncertainties in the TOAs were estimated from the least squares
procedure, and a minimum error, dependent on the observing system and
frequency, was required for each TOA.  In addition, a cut was
established such that all Mark~IV TOAs with least-squares
uncertainties greater than 10\,$\mu$s (roughly 20\% of the data) were
discarded; tests show that they do not significantly affect the
best-fit solution or the parameter uncertainties, and discarding these
data points allows us to better assess the data quality
(\S\ref{sec:confidence}).  The observatory's local time standard was
corrected retroactively to the UTC timescale, using data from the
Global Positioning System (GPS) satellites.

\section{Data Analysis}
\subsection{The Timing Model}\label{sec:model}

A pulse received on Earth at topocentric time $t$ is emitted at a time
in the comoving pulsar frame given by
\begin{equation}
T =  t-t_0+\Delta_C-D/f^2 + \Delta_{R\odot} + \Delta_{E\odot}
  -\Delta_{S\odot} - \Delta_R - \Delta_E - \Delta_S\,.
\label{eqn:orbit}
\end{equation}
Here $t_0$ is a reference epoch and $\Delta_C$ is the offset between
the observatory master clock and the reference standard of terrestrial
time.  The dispersive delay is $D/f^2$, where $D={\rm
DM}/2.41\times10^{-4}$, with the dispersion measure DM in cm$^{-3}$pc,
the radio frequency $f$ in MHz, and the delay in seconds.  The
$\Delta_{R\odot}$, $\Delta_{E\odot}$, and $\Delta_{S\odot}$ terms are
propagation delays and relativistic time adjustments for effects
within the solar system, and $\Delta_R$, $\Delta_E$ and $\Delta_S$ are
similar terms accounting for phenomena within the pulsar's orbit.

In Paper~I, we set out the equations that define the three orbital
terms $\Delta_R$, $\Delta_E$, and $\Delta_S$ in terms of ten
parameters.  Five are the standard Keplerian parameters: the period
$P_{b}$, projected semi-major axis $x\equiv a_1\sin i/c$, eccentricity
$e$, longitude of periastron $\omega$, and time of periastron $T_0$.
The other five are Post Keplerian (PK) parameters: the rate of advance
of the periastron $\dot\omega$, orbital period derivative $\dot
P_{b}$, time dilation and gravitational redshift $\gamma$, and range
$r$ and shape $s$ of the Shapiro time delay.  These quantities, in
conjunction with the time-variable dispersion measure, a simple time
polynomial to model the spin of the pulsar, and astrometric parameters
to model the propagation of the signal across the solar system,
constitute the free parameters to be fit in the theory-independent
timing model.

Within the framework of a particular theory of gravity, the five PK
parameters can be written as functions of the pulsar and companion
star masses, $m_1$ and $m_2$, and the well-determined Keplerian
parameters.  In general relativity the equations are as follows
\citep[see][]{dd86,tw89,dt92}:
\begin{eqnarray}
\dot\omega &=& 3 \left(\frac{P_b}{2\pi}\right)^{-5/3}
  (T_\odot M)^{2/3}\,(1-e^2)^{-1}\,, \label{eq:omdot} \\
\gamma &=& e \left(\frac{P_b}{2\pi}\right)^{1/3}
  T_\odot^{2/3}\,M^{-4/3}\,m_2\,(m_1+2m_2) \,, \\
\dot P_b &=& -\,\frac{192\pi}{5}
  \left(\frac{P_b}{2\pi}\right)^{-5/3}
  \left(1 + \frac{73}{24} e^2 + \frac{37}{96} e^4 \right)
  (1-e^2)^{-7/2}\,T_\odot^{5/3}\, m_1\, m_2\, M^{-1/3}\,,
  \label{eq:pbdot} \\
r &=& T_\odot\, m_2\,,\label{eq:r} \\
s &=& x \left(\frac{P_b}{2\pi}\right)^{-2/3}
  T_\odot^{-1/3}\,M^{2/3}\,m_2^{-1}\,. \label{eq:s}
\end{eqnarray}
Here the masses $m_1$, $m_2$, and $M\equiv m_1+m_2$ are expressed in
solar units, $T_\odot\equiv GM_\odot/c^3 = 4.925490947\,\mu$s, $G$ the
Newtonian constant of gravity, and $c$ the speed of light.

\subsection{Arrival Time Analysis}\label{sec:timing}

\begin{figure}
\plotone{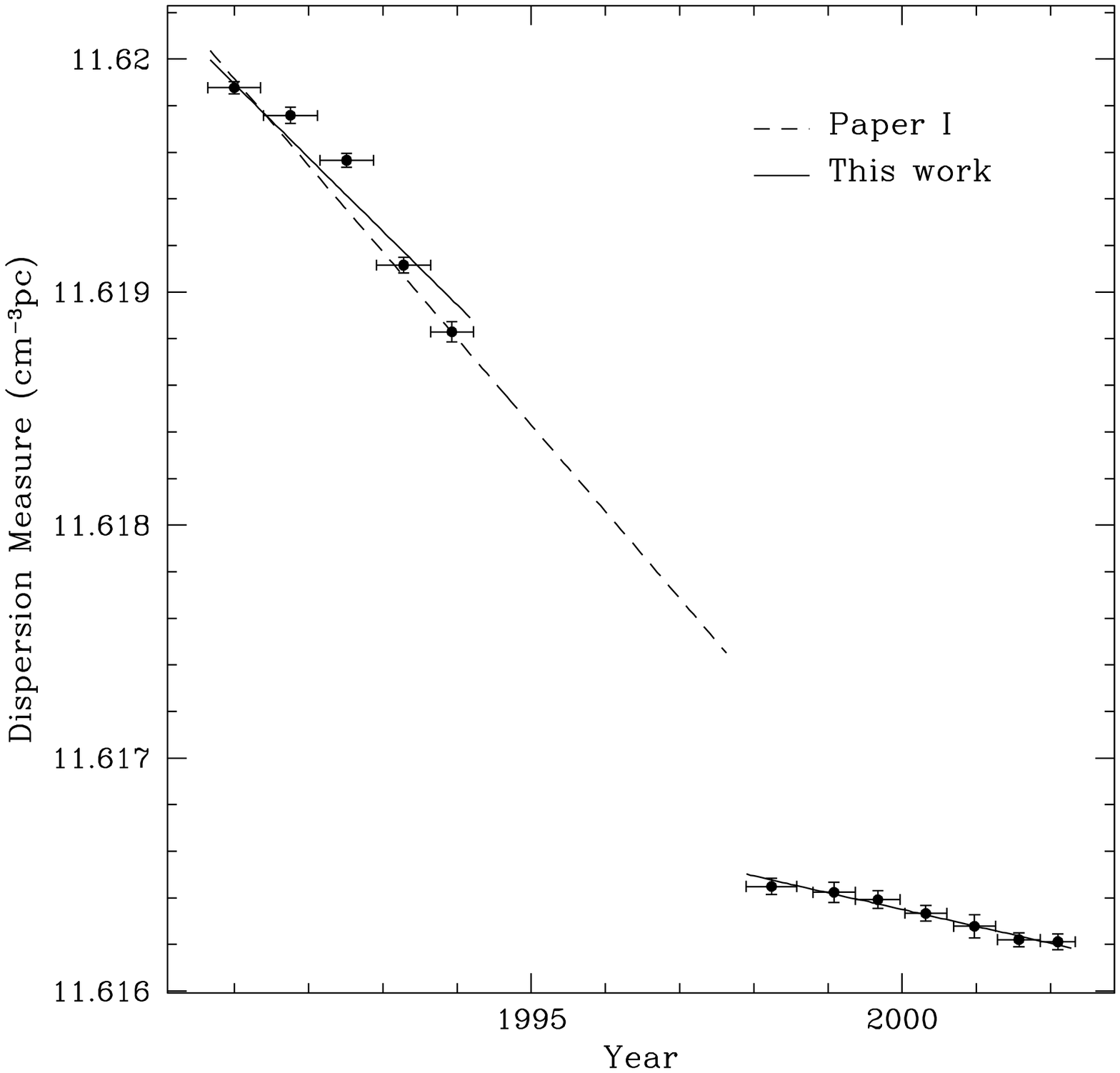}
\figcaption{The dispersion measure (DM) used in the fits presented
in Tables~\ref{tab:astspin} and \ref{tab:orbparms}.  Different values
and derivatives are needed for the pre- and post-upgrade data sets.
The DM and its variation obtained in Paper~I are shown for comparison;
they match our current results within the uncertainties.  The points
with error bars illustrate the quality of DM fits over short
intervals: the horizontal errors bars indicate the domain of each fit,
while the vertical error bars represent the uncertainties derived in
each time bin.
\label{fig:dmvar}}
\end{figure}

We used the standard {\sc tempo} analysis software \citep[see also
\url{http://pulsar.princeton.edu/tempo}]{tw89}
together with the JPL~DE200 solar-system ephemeris \citep{sta90} to
perform a least-squares fit of the measured pulse arrival times to the
timing model.  Results for the astrometric, spin, and dispersion
parameters of PSR B1534+12 are presented in Table~\ref{tab:astspin}.
The DM was fit in two blocks, pre-and post-upgrade, with different
base values and time derivatives in each block.  Because of systematic
errors (see \S\ref{sec:confidence} below), the Mark~III 430\,MHz data
were only used to fit the pre-upgrade DM and DM derivative, and were
not incorporated into the rest of the timing solution.  This meant
that an iterative approach was required to find the best-fit DMs:
first the spin and orbital solution was held constant while the DM
parameters were fit using the whole dataset, then the Mark~III
430\,MHz data were dropped as the best overall timing solution was
obtained from the other data, holding the pre-upgrade DM parameters
fixed.  This new timing solution was then used to refine both sets of
DM parameters, until no significant changes resulted.  The DM and
DM-derivative uncertainties reported in Table~\ref{tab:astspin} are
the errors derived while holding the rest of the timing solution
fixed.  Fitting for post-upgrade DM and DM-derivative simultaneously
with the rest of the timing solution yielded similar values, but
somewhat larger uncertainties due to covariances with other
parameters.  Figure~\ref{fig:dmvar} shows the time variation of the
best-fit DM, with the DM and first derivative from Paper~I shown for
comparison; the pre-upgrade values are consistent within the
uncertainties.  We note that the offset in DM base values between the
pre- and post-upgrade fits is due mostly to the use of different
standard profiles with necessarily different frequency alignments;
thus this offset should not be interpreted as physically meaningful.
The change in DM {\it derivative}, however, is real, and is discussed
in \S\ref{sec:dmvar} below.  An arbitrary time offset was also allowed
between the pre- and post-upgrade data sets to accommodate the
different standard profiles and changes in the instrumental signal
path length.

Two models of the pulsar orbit were used to fit the data.  The
theory-independent ``DD'' model \citep{dd86} treats all five PK
parameters defined in \S\ref{sec:model} as free parameters in the fit.
The alternate ``DDGR'' model \citep{tay87a,tw89} assumes that general
relativity is correct and uses equations~\ref{eq:omdot}
through~\ref{eq:s} to tie the PK parameters to $M\equiv m_1+m_2$ and
$m_2$; consequently it requires only two post-Keplerian free
parameters.

Table~\ref{tab:orbparms} presents our best-fit orbital parameters.
Uncertainties given in the table are approximately twice the formal
``$1\,\sigma$'' errors from the fit; we believe them to be
conservative estimates of the true 68\%-confidence uncertainties,
including both random and systematic effects.  For each model, we list
the Keplerian parameters and the relevant post-Keplerian parameters.
We also include, in italic numbers, the computed PK parameters derived
from the measured masses in the DDGR fit; these computed parameters
are in excellent agreement with the DD fit values.  As in Paper~I, the
DDGR solution includes the ``excess $\dot P_b$'' parameter, which
accounts for an otherwise unmodeled acceleration resulting from
galactic kinematics.  We note that the measured orbital period
derivative, $-0.137\pm0.003 \times 10^{-12}$, is very close to the sum
of the DDGR model's intrinsic value, $-0.1924\times 10^{-12}$, and the
fitted excess, $+0.055\pm0.003 \times 10^{-12}$.

\begin{figure}
\plotone{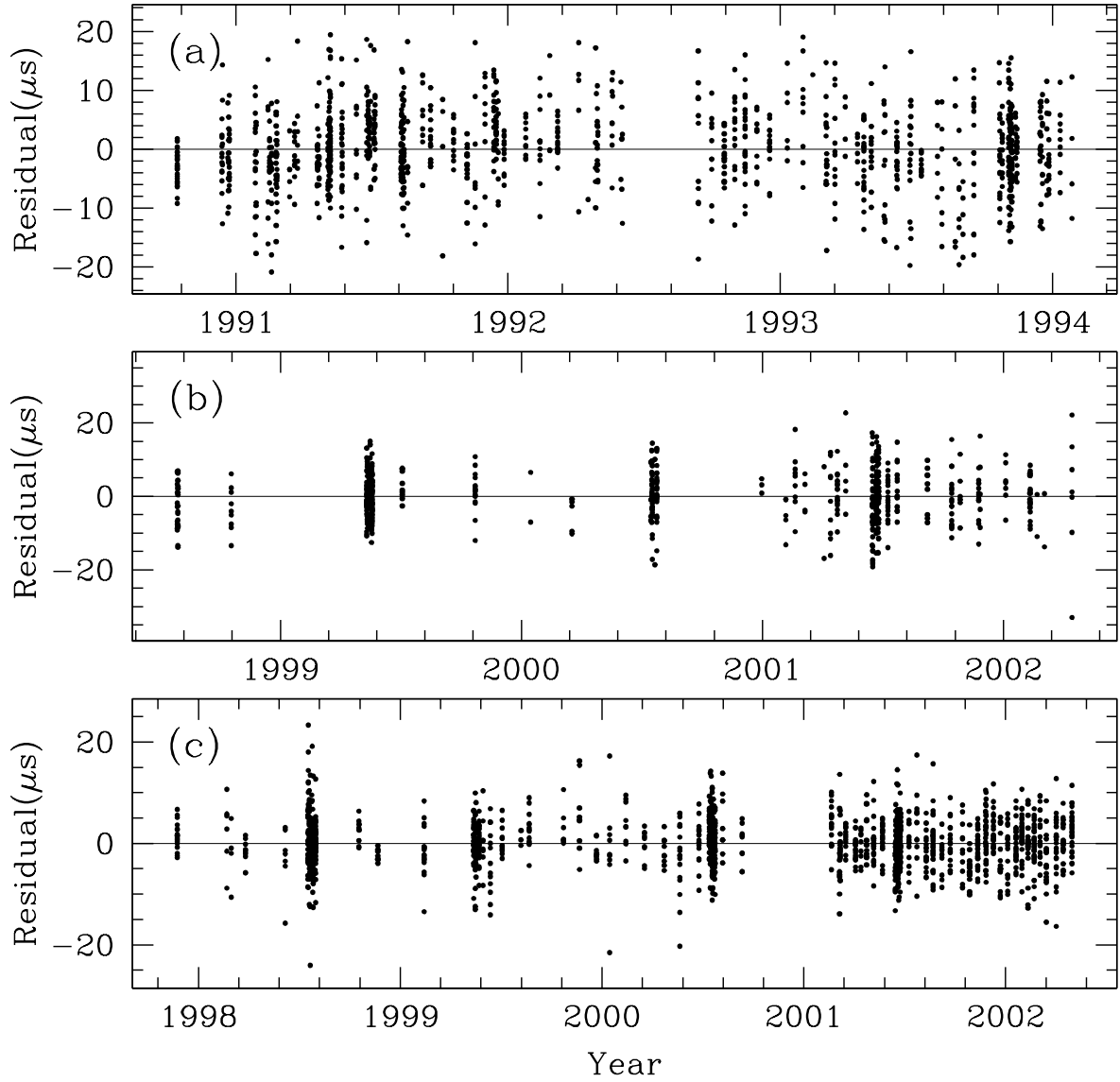}
\figcaption{Post-fit residuals versus date for 
(a) the pre-upgrade (Mark III) 1400\,MHz data, 
(b) the post-upgrade (Mark IV) 1400\,MHz data, and
(c) the post-upgrade (Mark IV) 430\,MHz data. 
\label{fig:dayres}}
\end{figure}

\begin{figure}
\plotone{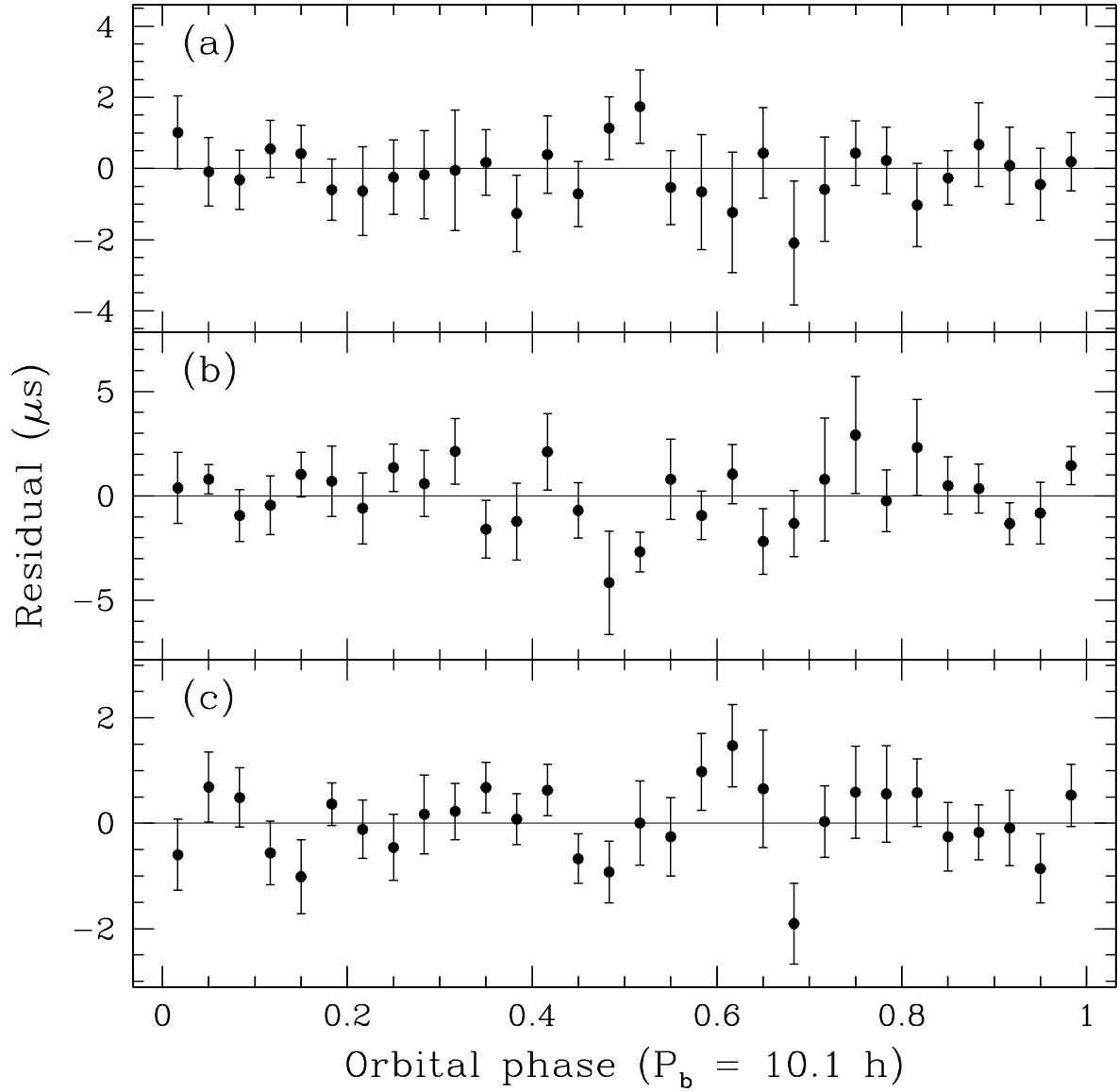}
\figcaption{Average post-fit residuals as a function of
orbital phase for (a) the pre-upgrade (Mark III) 1400\,MHz data, 
(b) the post-upgrade (Mark IV) 1400\,MHz data, and
(c) the post-upgrade (Mark IV) 430\,MHz data. \label{fig:orbbin}}
\end{figure}

Figure~\ref{fig:dayres} shows the post-fit residuals for the
pre-upgrade 1400\,MHz data, and the post-upgrade 1400 and 430\,MHz
data, plotted as functions of date.  Interstellar scintillation causes
the TOA uncertainties to vary somewhat, but in the interest of clarity
we have not drawn error bars on these residual plots.
Figure~\ref{fig:orbbin} illustrates the averaged post-fit residuals
for the same datasets, plotted as functions of orbital phase.

\subsection{Comparison of Coherent and Incoherent Observing Systems}\label{sec:confidence}

As discussed in Paper~I, solutions for the fitted parameters of
PSR~B1534+12, as determined from the pre-upgrade 430\,MHz data, are
biased by small systematic errors in the TOAs, likely caused by
imperfect post-detection dispersion removal combined with variable
interstellar scintillation patterns.  These measurements are
unreliable at the several microsecond level required for
high-precision timing of this pulsar.

\begin{figure}
\plotone{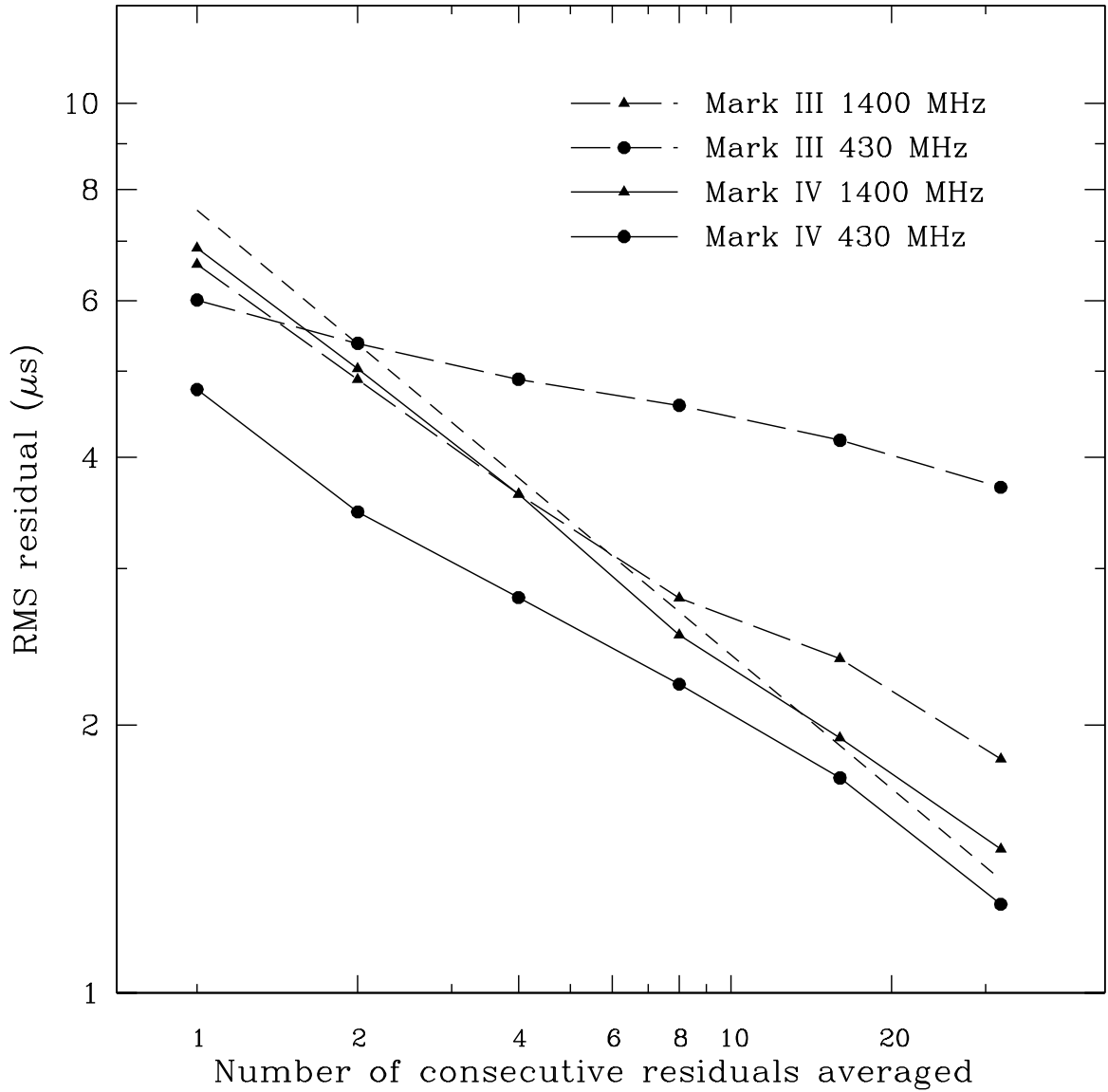}
\figcaption{Root-mean-square residual versus number of
consecutive residuals averaged, for the four data sets: pre-upgrade
(Mark~III) and post-upgrade (Mark~IV) data at each of 1400 and
430\,MHz.  The short-dashed line indicates the expected slope of $-1/2$ for
uncorrelated residuals.  All data sets except the pre-upgrade 430\,MHz
data are in reasonable agreement with this prediction.
\label{fig:resn}}
\end{figure}

One of our major goals in building the Princeton Mark~IV instrument
\citep{sst+00} was to improve the timing accuracy at low frequencies 
for fast systems such as PSR~B1534+12, by minimizing any
dispersion-related systematic errors in TOAs.  We have determined that
the new instrument does indeed appear to have circumvented the
residual dispersion problems that plague the Mark~III data at low
frequencies.  In Figure~\ref{fig:resn} we follow \citet{tw89} and
Paper~I in examining the statistical properties of the post-fit
residuals (relative to the solution of Tables~\ref{tab:astspin} and
\ref{tab:orbparms}) to see if they ``integrate down'' as $n^{-1/2}$
when $n$ consecutive values are averaged.  The pre-upgrade 1400\,MHz
data and both post-upgrade data sets are fairly consistent with the
expected slope, while the pre-upgrade 430\,MHz data deviate
significantly.  This confirms that we have been justified in leaving
the pre-upgrade 430\,MHz data out of our preferred timing solution,
that these Mark~III 430\,MHz data indeed contain systematic errors
related to the incoherent dispersion removal, and that the new Mark~IV
instrument greatly reduces these systematics, as we had hoped.  The
minimum TOA uncertainty requirement mentioned above was introduced to
compensate for the small remaining deviations from the $n^{-1/2}$ law
visible in the three good datasets.

\section{Discussion} \label{sec:disc}

\subsection{Observed Change in Orbital Period:  Distance to PSR\,B1534+12}

The observed $\dot P_b$ is measured in the reference frame of the
solar system barycenter, and must be corrected to the center-of-mass
frame of the binary pulsar system before it can be compared to the
value predicted by GR.  The largest correction is kinematic, involving
the relative acceleration of the two reference frames.  It can be
broken down into the vertical acceleration in the Galactic potential,
acceleration in the plane of the Galaxy, and an apparent centripetal
acceleration due to the transverse velocity of the pulsar binary
\citep{dt91}:
\begin{equation}\label{eq:gal}
\left(\frac{\dot{P_b}}{P_b}\right)^{\rm gal} = -\,\frac{a_z\sin b}{c}
 \,-\,\frac{v_0^2}{cR_0} \left[\cos l + 
  \frac{\beta}{\sin^2 l + \beta^2}\right] +\mu^2\frac{d}{c}.
\end{equation}
Here $a_z$ is the vertical component of galactic acceleration, $l$ and
$b$ the pulsar's galactic coordinates, $R_0$ and $v_0$ the Sun's
galactocentric distance and galactic circular velocity, $\mu$ and $d$
the pulsar's proper motion and distance, and $\beta=d/R_0 - \cos l$.
As we do not have a precise distance to the pulsar through either a
timing or an interferometric parallax measurement, the best
independent distance estimate still comes from the \citet{tc93} model
of the free electron content of the Galaxy; this model puts the pulsar
at a distance of $0.7\pm0.2$\,kpc.  At this distance the \citet{kg89}
model of the Galactic potential yields an estimate of
$a_z/c=(1.60\pm0.13)\times10^{-19}\,\mbox{s}^{-1}$.  We assume
$v_0=222\pm20\,$km\,s$^{-1}$ and $R_0 = 7.7\pm0.7$\,kpc, as in
\citet{dt91}.  Then, summing the terms in equation~(\ref{eq:gal}) and
multiplying by $P_b$, we find the total kinematic bias to be
\begin{equation}
\left(\dot P_b\right)^{\rm gal} = (0.037\pm0.011)\times10^{-12}\,.
\end{equation}
The uncertainty in this correction is dominated by the uncertainty in
distance, which is only roughly estimated by the Taylor and Cordes
model.  The slight decrease in the uncertainty from that given in
Paper~I results from an increase by a factor of ten in the precision
of the proper motion measurement.

Our measurement of the intrinsic rate of orbital period decay is
therefore
\begin{equation}
\left(\dot P_b\right)^{\rm obs} - \left(\dot P_b\right)^{\rm gal}
        = (-0.174\pm0.011)\times 10^{-12}\,.
\end{equation}
The uncertainty is completely dominated by the uncertainty on the
kinematic correction, which is nearly a factor of 4 larger than the
measurement uncertainty on $\dot P_b^{\rm obs}$.  In GR, the orbital
period decay due to gravitational radiation damping, $(\dot{P_b})^{\rm
GR}$, can be predicted from the masses $m_1$ and $m_2$
(eq.~\ref{eq:pbdot}), which in turn can be deduced from the high
precision measurements of $\dot{\omega}$ and $\gamma$.  As listed in
Table~\ref{tab:orbparms}, the expected value is
\begin{equation}
\left(\dot P_b\right)^{\rm GR} = -0.192\times 10^{-12}\,.\label{eq:pbdotgr}
\end{equation}
As in Paper~I, our measured value differs from this prediction, now by
some 1.7 standard deviations, and again, assuming that GR is the
correct theory of gravity, we can derive the true distance to the
pulsar from the ``excess $\dot P_b$'' parameter and
equation~\ref{eq:gal} above \citep{bb96}.  Our improved distance is
$d=1.02\pm0.05$~kpc (68\% confidence limit).  The uncertainty is still
dominated by the measurement uncertainty of $(\dot P_b)^{\rm obs}$,
rather than uncertainties in the galactic rotation parameters or the
acceleration $a_z$, though the Galactic model uncertainties will
ultimately limit the distance measurement to a few percent accuracy.
The timing parallax for this system is still not significantly
measured, but is constrained to be less than 1.5\,mas, in good
agreement with the GR-derived result.

Our new distance result is consistent with the $1.1\pm0.2$\,kpc
determined in Paper~I, but we have improved the uncertainty by a
factor of 4.  Our previous distance led to a downward revision of the
estimated double-neutron-star inspiral rate visible to
gravitational-wave observatories such as LIGO.  The new measurement
leads to a small (15\%) increase in the number density of similar
systems in the local universe, with the uncertainty now dominated by
the uncertain scale height of such binaries \citep[e.g.,][]{knst01}.

\subsection{Dispersion Measurements and the Local Interstellar Medium}\label{sec:dmvar}

As noted in \S\ref{sec:timing}, we find a significantly different DM
and DM derivative in the post-upgrade era than before the upgrade
(Figure~\ref{fig:dmvar}).  We argue that while the change in reference
DM value is largely due to the use of new standard profiles with a
slightly different frequency alignment, the difference in the rate of
change (more than a factor of 4 slower since the upgrade) is physical.
It is true that the profile at both frequencies is undergoing a
secular evolution due to geodetic precession of the pulsar's spin axis
\citep[][Stairs {\it et al.}, in prep.]{arz95} and thus one might
suspect a contribution to the DM derivative from different temporal
evolution at the two observing frequencies.  We explore this
possibility by calculating the DM difference between our 1999 May and
2001 June observing campaigns, for the best-fit post-upgrade DM
derivative versus the best-fit {\it pre-}upgrade DM derivative,
finding that if the pre-upgrade DM trend had continued, the 2001 June
DM would have been smaller by about 0.0005 cm$^{-3}$\,pc.  This large
a difference would imply a difference in offset between the 430\,MHz
and 1400\,MHz profiles of 10\,$\mu$s, or 0.27 bins in a 1024-bin
profile.  By comparing the actual cumulative profiles at these epochs
using the same cross-correlation algorithm used to calculate TOAs, we
estimate that the 430--1400\,MHz offset has shifted by no more than
about 0.05 bins.  As the TOAs are determined by the strong main peak
of the pulse and the most noticeable evolution is taking place in the
low-level emission near the base of the pulse \citep{stta00} this is
perhaps not surprising.  We therefore conclude that the observed
changes in DM slope do represent changes in the structure of the
interstellar medium between the Earth and the pulsar.

With our current data set, it is not possible to construct a
satisfactory structure function with which to characterize the length
scales of turbulence in the interstellar medium \citep[e.g.,][]{pw91}.
Because of strong refractive scintillation at 1400\,MHz, the
high-frequency data set remains quite sparse
(Figure~\ref{fig:dayres}). Even though we may use a wider-bandwidth
instrument such as the recently commissioned 100-MHz ``WAPP''
spectrometer for future 1400\,MHz observations, the broadband nature
of the refractive interstellar scintillation may make it difficult to
accurately sample the dispersion measure variations with resolution
finer than roughly 6 months.

\subsection{The Neutron Star Masses: Testing Evolutionary Theory}

\begin{figure}
\plotone{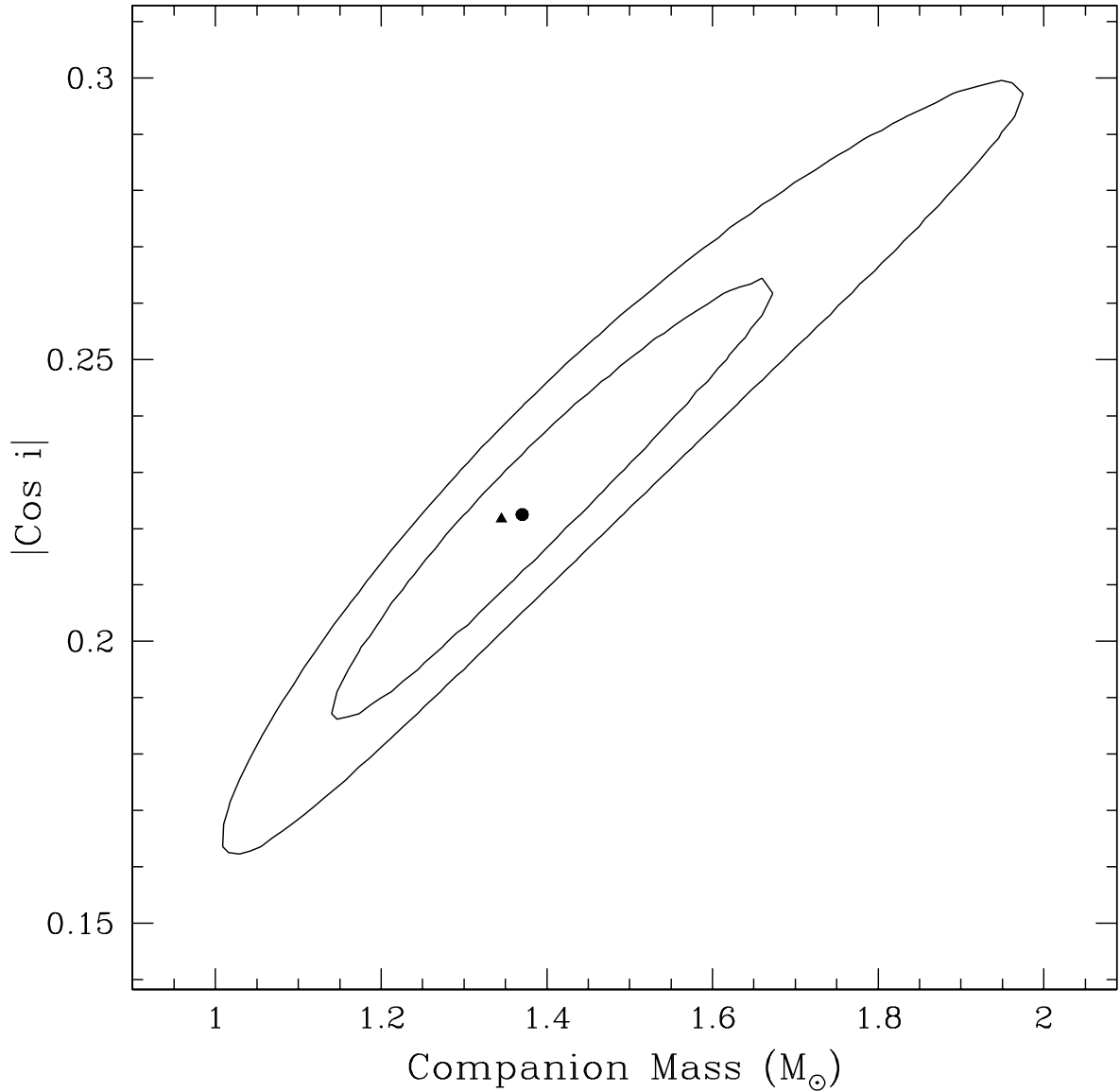}
\figcaption{The Shapiro delay parameters.  Contours indicate the 68\% 
and 95\% confidence levels for the companion mass and the cosine of
the orbital inclination.  The round dot represents the best-fit DD
solution in Table~\ref{tab:orbparms} while the triangle represents the
optimal DDGR solution, which is well within the 68\% contour of
the DD solution.
\label{fig:chisq}}
\end{figure}

The best mass estimates for the two neutron stars in the PSR~B1534+12
system come from the DDGR timing model.  The measured values are $m_1
= 1.3332\pm0.0010\,M_{\odot}$ and $m_2 = 1.3452\pm0.0010\,M_{\odot}$.
The masses derived from the DD model and equations~\ref{eq:r} and
\ref{eq:s} are in good agreement with these values, though the 
measurement uncertainties are much larger (Figure~\ref{fig:chisq}).

Although the masses of the two neutron stars are very similar, it is
now clear that the pulsar is significantly less massive than its
companion.  This is contrary to initial expectations from binary
evolution.  The pulsar is rapidly spinning, and has the low magnetic
field characteristic of a ``recycled pulsar'' that has been spun-up by
mass transfer from a companion star.  Assuming a monotonic relation
between progenitor mass and neutron star mass, we therefore might
predict that the pulsar would be more massive at birth, and that the
mass difference would only increase during the later evolution of the
system.

This perhaps naive prediction has already been challenged by the
binary pulsar system PSR~B2303+46. Recently, the companion in this
close, eccentric binary was shown to be a white dwarf rather than a
second neutron star \citep{vk99}.  The eccentric orbit implies that
the white dwarf formed first, from the star that was originally more
massive.  Presumably, mass transfer to the initially less massive star
pushed it above the minimum mass needed to form a supernova and a
neutron star.

A similar process may have been important in the formation of the
PSR~B1534+12 system.  Although the pulsar was formed from the
initially more massive star, mass transfer to its companion star
probably resulted in a mass inversion before either neutron star was
formed.

To move beyond the simplest (or most naive) predictions about the
relative masses of the pulsar and companion in the B1534+12 system
requires significant improvements in our theoretical understanding of
several areas of stellar and binary evolution.  First, we must
understand the relationship between progenitor and neutron star
masses, which may not be purely monotonic.  Then we must understand
how the mass lost by the pulsar's progenitor or gained by the
companion's progenitor affected the evolution of the stellar cores and
the amount of fallback during the supernova events.  Finally, we must
understand in some detail the mass transfer that spun up the pulsar.
Although these are all challenging problems, there is some hope for
future guidance from the growing number of precisely measured neutron
star masses.

\subsection{Test of Relativity}

\begin{figure}
\plotone{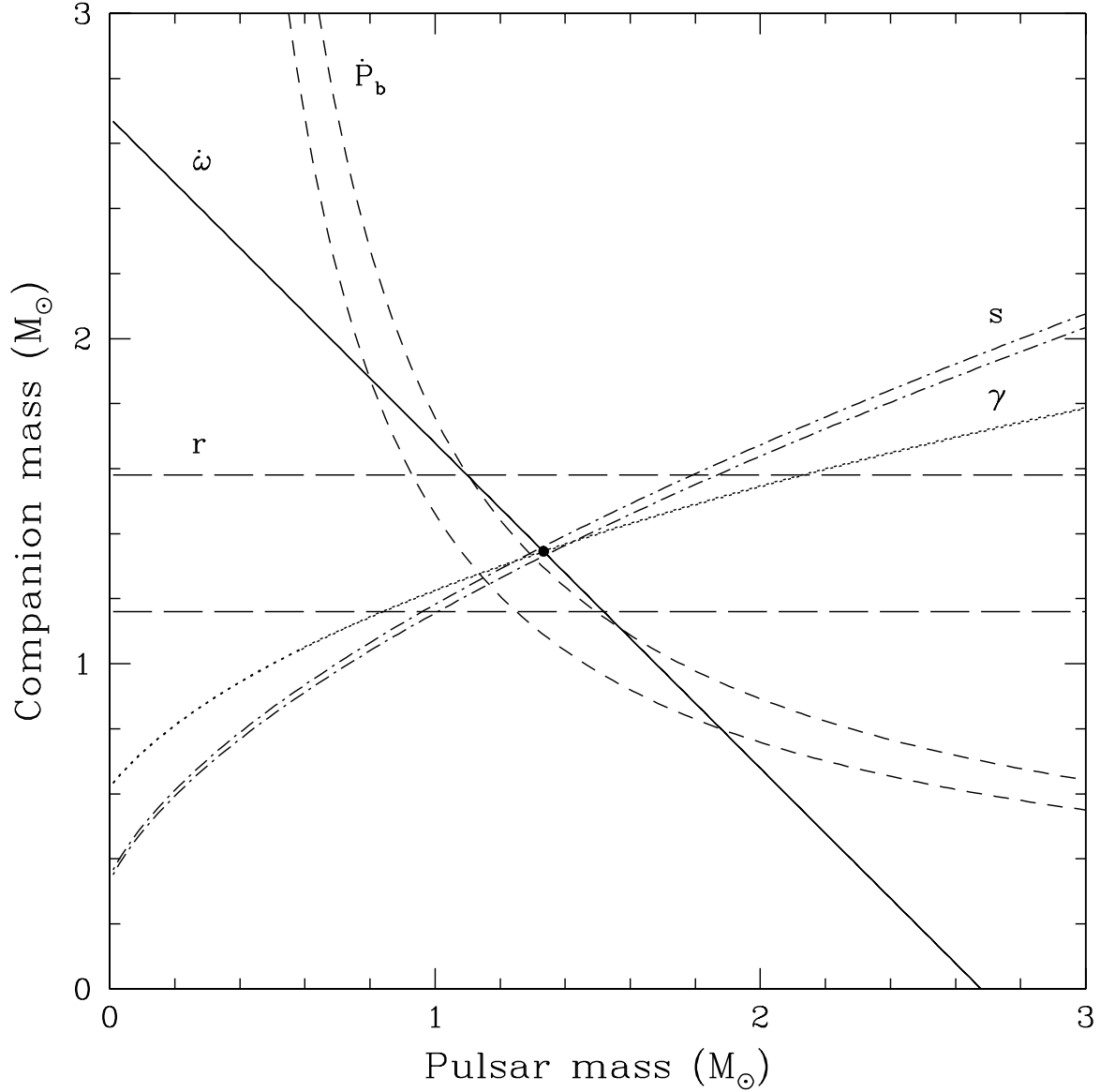}
\figcaption{Mass-mass diagram for the PSR B1534+12
system.  Labeled curves illustrate 68\% confidence ranges of the DD
parameters listed in Table~\ref{tab:orbparms}.  The filled circle
denotes the component masses according to the DDGR solution.  A
kinematic correction for assumed distance $d=0.7\pm0.2\,$kpc has been
subtracted from the observed value of $\dot{P_b}$; the uncertainty on
this kinematic correction dominates the uncertainty of this curve.  A
slightly larger distance removes the small apparent discrepancy
between the observed and predicted values of
$\dot{P_b}$. \label{fig:massmass}}
\end{figure}

PSR~B1534+12 permits the second test of general relativity based on
the $\dot\omega$, $\gamma$, and $\dot P_b$ parameters of a binary
pulsar system, and adds significant measurements of the Shapiro-delay
parameters $r$ and $s$.  The left-hand sides of
equations~(\ref{eq:omdot}--\ref{eq:s}) represent quantities measured
in a theory-independent fashion and listed in the DD column of
Table~\ref{tab:orbparms}.  If GR is consistent with these measurements
and there are no significant unmodeled effects, the five curves
corresponding to equations~(\ref{eq:omdot}--\ref{eq:s}) should
intersect at a single point in the $m_1$-$m_2$ plane.  These curves
are presented in Figure~\ref{fig:massmass}, in which a pair of lines
delimit the 68\% confidence limit for each PK parameter (a single line
is drawn for $\dot\omega$, whose uncertainty is too small to show).
It is clear that the $\dot\omega$, $\gamma$, $r$, and $s$ curves
intersect, though $r$ is still poorly measured.  The curve obtained
from the observed DD value of $\dot P_b$ can be made to intersect the
others, as discussed above, by setting the pulsar distance to
$1.02\pm0.05$\,kpc rather than the $0.7\pm0.2$\,kpc estimated from the
dispersion measure.  A filled circle at $m_1=1.3332~M_\odot,
m_2=1.3452~M_\odot$ marks the DDGR solution of
Table~\ref{tab:orbparms}, and its location on the $\dot\omega$ line
agrees to within 0.05\% with the measured DD values of $\gamma$ and
$s$.  This provides a highly precise test of the validity of general
relativity using only non-radiative timing parameters, an important
complement to the mixed $\dot\omega$--$\gamma$--$\dot P_b$ test
provided by PSR~B1913+16.

\acknowledgements

The Arecibo Observatory, a facility of the National Astronomy and
Ionosphere Center, is operated by Cornell University under a
cooperative agreement with the National Science Foundation.  We thank
Zaven Arzoumanian, Fernando Camilo, Andrew Lyne and David Nice for
their earlier contributions to this long-term project, Kiriaki
Xilouris, Duncan Lorimer, David Nice, Eric Splaver, Andrea Lommen,
Paulo Freire, and Ian Hoffman for assistance with observations, and an
anonymous referee for comments on the manuscript..  I.~H.~S. is a
Jansky Fellow.  S.~E.~T. is supported by the NSF under grant
AST-0098343.  J.~H.~T. is supported by the NSF under grant
AST96-18357.  A.~W. is supported by the NSF under grant AST-9988217.

\clearpage


\clearpage

\clearpage

\begin{deluxetable}{lrrrr}
\tablecolumns{5}
\tablewidth{0pt}
\tablecaption{Parameters of the four observing systems.\label{tab:obsparms}}
\tablehead{\colhead{} & \colhead{Mark III} & \colhead{Mark III}
 & \colhead{Mark IV} & \colhead{Mark IV}}
\startdata
Frequency (MHz) \dotfill  & 1400 & 430 & 1400 & 430 \\
Bandwidth (MHz) \dotfill  & 40 & 8 & 5 & 5 \\
Spectral Channels \dotfill  & 32 & 32 & 1 & 1 \\
Dedispersing system \dotfill & incoherent & incoherent 
  & coherent & coherent \\
Time resolution ($\mu$s) \dots  & 97 & 329 & 38 & 38 \\
Integration time (s) \dotfill  & 300 & 180 & 190 & 190 \\
Dates \dotfill &1990.8--94.1 &1990.7--94.2 & 1998.6--2002.3 & 1997.9--2002.3 \\
Number of TOAs \dotfill  & 1185 & 2311 & 639 & 1635 \\
Median $\sigma_{\rm TOA}$ ($\mu$s) \dotfill  & 6.1  & 6.0 & 6.7 & 4.2 \\
\enddata
\end{deluxetable}
 
\begin{deluxetable}{ll}
\tablecolumns{2}
\tablewidth{0pt}
\tablecaption{Astrometric, spin, and dispersion
parameters for PSR B1534+12\tablenotemark{a}.\label{tab:astspin}}
\tablehead{\colhead{Parameter} & \colhead{Value}}
\startdata
Right ascension, $\alpha$ (J2000)\dotfill  & 
  $15^{\rm h}\,37^{\rm m}\,09\fs960312(10)$ \\
Declination, $\delta$ (J2000)  \dotfill  & 
  $11^\circ\,55'\,55\farcs5543(2)$ \\
Proper motion in R.A., $\mu_\alpha$ (mas\,yr$^{-1}$)  \dotfill  & 
1.32(3) \\
Proper motion in Dec., $\mu_\delta$ (mas\,yr$^{-1}$)  \dotfill  & 
$-$25.12(5) \\
Parallax, $\pi$ (mas) \dotfill & $<1.5$ \\
\\
Pulse period, $P$ (ms)  \dotfill  & 
37.9044407982695(4) \\
Period derivative, $\dot P$ $(10^{-18})$ \dotfill & 2.422622(3) \\
Epoch (MJD) \dotfill & 50261.0 \\
\\
Dispersion measure, DM, 1 (cm$^{-3}$pc) \dotfill   & 
11.61944(2) \\
DM derivative 1 (cm$^{-3}\mbox{pc\,yr}^{-1}$)  \dotfill  & 
$-$0.000316(10) \\
DM Epoch 1 (MJD) \dotfill   &  48778.0 \\
\\
DM 2 (cm$^{-3}$pc) \dotfill   & 
11.61634(3) \\
DM derivative 2 (cm$^{-3}\mbox{pc\,yr}^{-1}$)  \dotfill  & 
$-$0.000073(7) \\
DM Epoch 2 (MJD) \dotfill   &  51585.0 \\
\\
Galactic longitude $l$ (deg) \dotfill  & {\it 19.8} \\ 
Galactic latitude $b$ (deg) \dotfill  & {\it 48.3} \\
Composite proper motion, $\mu$ (mas\,yr$^{-1}$) \dotfill & {\it 25.15(5)} \\
Galactic position angle of $\mu$ (deg) \dotfill  & {\it 239.16(8)} \\
\enddata
\tablenotetext{a}{Figures in parentheses are uncertainties in the last digits
quoted, and italic numbers represent derived quantities.}
\end{deluxetable}

\begin{deluxetable}{lll}
\tablewidth{0pt}
\tablecolumns{3}
\tablecaption{Orbital parameters of PSR B1534+12 in the DD and DDGR models\tablenotemark{a}.\label{tab:orbparms}}
\tablehead{\colhead{} &\colhead{ DD model} & \colhead{DDGR model}}
\startdata
Orbital period, $P_b$ (d) \dotfill  & 0.420737299122(10) & 0.420737299123(10) \\ 
Projected semi-major axis, $x$ (s) \dotfill  & 3.729464(2) & 3.7294641(4) \\
Eccentricity, $e$ \dotfill  & 0.2736775(3) & 0.27367740(14) \\ 
Longitude of periastron, $\omega$ (deg) \dotfill  & 274.57679(5) 
  & 274.57680(4) \\
Epoch of periastron, $T_0$ (MJD) \dotfill  & 50260.92493075(4) 
  & 50260.92493075(4) \\
\\
Advance of periastron, $\dot\omega$ (deg\,yr$^{-1}$) \dotfill  
  & 1.755789(9) & {\it 1.7557896} \\
Gravitational redshift, $\gamma$ (ms) \dotfill  & 2.070(2) & {\it 2.069} \\
Orbital period derivative, $(\dot P_b)^{\rm obs}$ $(10^{-12})$ \dotfill  &
  $-$0.137(3) & {\it $-$0.1924} \\
Shape of Shapiro delay, $s$ \dotfill  & 0.975(7) & {\it 0.9751} \\
Range of Shapiro delay, $r$ ($\mu{\rm s}$) \dotfill  & 6.7(1.0) & 
 {\it 6.626}  \\
Derivative of $x$, $\left |\dot x \right |$ $(10^{-12})$ \dotfill & $<0.68$
& $<0.015$ \\
Derivative of $e$, $\left |\dot e \right |$ $(10^{-15}\,{\rm s}^{-1})$
\dotfill & $<3$ & $<3$ \\
\\
Total mass, $M=m_1+m_2$ ($M_{\odot}$) \dotfill  & \nodata & 2.678428(18) \\
Companion mass, $m_2$ ($M_{\odot}$) \dotfill    & \nodata & 1.3452(10) \\
Excess $\dot P_b$ $(10^{-12})$ \dotfill         & \nodata & 0.055(3) \\
\enddata
\tablenotetext{a}{Figures in parentheses are uncertainties in the last digits
quoted.  Italic numbers represent derived parameters, assuming
general relativity.}
\end{deluxetable}

\end{document}